# Multipole Engineering of Attractive-Repulsive and Bending Optical Forces

Denis A. Kislov[1], Egor A. Gurvitz[1], Alexander A. Pavlov[2], Dmitrii N. Redka[3], Manuel I. Marqués[4], Pavel Ginzburg[5,6] and Alexander S. Shalin[1,2,7]

[1] ITMO University Kronverksky prospect 49, St. Petersburg 197101, Russia
[2] Institute of Nanotechnology of Microelectronics of the Russian Academy of Sciences, Moscow 119991, Russia
[3] Saint Petersburg Electrotechnical University "LETI" (ETU), 5 Prof. Popova Street, 197376, St. Petersburg, Russia
[4] Departamento de Física de Materiales, IFIMAC and Instituto "Nicolás Cabrera", Universidad Autónoma de Madrid, 28049 Madrid, Spain
[5] School of Electrical Engineering, Tel Aviv University, Tel Aviv 69978, Israel
[6] Center for Photonics and 2D Materials, Moscow Institute of Physics and Technology, Dolgoprudny 141700, Russia
[7] Kotel'nikov Institute of Radio Engineering and Electronics of Russian Academy of Sciences (Ulyanovsk branch), Goncharova Str.48, Ulyanovsk, Russia, 432000

**Abstract**:

Focused laser beams allow controlling mechanical motion of objects and can serve as a tool for assembling complex micro and nano structures in space. While in a vast majority of cases small particles experience attractive gradient forces and repulsive radiation pressure, introduction of additional degrees of freedom into optomechanical manipulation suggests approaching new capabilities. Here we analyze optical forces acting on a high refractive index silicon sphere in a focused Gaussian beam and reveal new regimes of particle's anti-trapping. Multipolar analysis allows separating an optical force into interception and recoil components, which have a completely different physical nature resulting in different mechanical actions. In particular, interplaying interception radial forces and multipolar resonances within a particle can lead to either trapping or anti-trapping scenarios, depending of the overall system parameters. At the same time, the recoil force generates a significant azimuthal component along with an angular-dependent radial force. Those contribution enable enhancing either trapping or anti-trapping regimes and also introduce bending reactions. These effects are linked to the far-field multipole interference resulting and, specifically, to its asymmetric scattering diagrams. The later approach is extremely useful, as it allows assessing the nature of optomechanical motion by observing far-field patterns. Multipolar engineering of optical forces, being quite general approach, is not necessarily linked to simple spherical shapes and paves a way to new possibilities in microfluidic applications, including sorting and micro assembly of nontrivial volumetric geometries.

*Corresponding Author: denis.a.kislov@gmail.com

# Introduction

Optomechanical manipulation, been first demonstrated by A. Ashkin at the middle of 1980s [1], opened numerous venues in fundamental and applied science, e.g. [2],[3],[4]. Classical configurations of optical tweezers include a high numerical aperture objective, which focuses a laser beam into a small, yet diffraction limited spot. Typically, zero-order Gaussian beams are the preferable choice for achieving a stable trapping. In the case of subwavelength particles, the main contributing terms are the gradient force and radiation pressure – finding a balance between those two enables immobilizing an object. While this classical configuration has been widely explored and used nowadays, introducing new degrees of freedom in optomechanical manipulation is the subject of intensive research. Those investigations are partially inspired by new microfluidic applications, where fast sorting [5]–[7] and mixing [8] of colloidal substances is one of the essential functions to have. Optomechanical tools are also frequently used in biological and biomedical investigations, where noninvasive *in vivo* manipulations are done with tissue-penetrating laser beams [9]–[11]. Furthermore, light-assisted targeted drug delivery (yet *in vitro*) [12], [13] and biosensing [14] are among other areas, where a flexible optomechanical manipulation can find a use.

Enlarging a number of optomechanical degrees of freedom can be obtained with three fundamentally different approaches, at least the main reports in the field can be classified by the following logic. The first method is based on shaping a laser beam. One of the main experimental techniques here is to use holographic masks, either static [15], [16] or reconfigurable [17], [18] (spatial light modulators are typically employed in the latter case). Holographic optical tweezers are used to trap multiple particles simultaneously [19]. Holographic masks are also used to generate non-Gaussian beams for optical trapping, e.g. Bessel beams [20], [21] and beams with inherent orbital angular momentum [22], [23].

Another approach to a flexible manipulation is to introduce auxiliary photonic structures, which assist configuring optical forces. Being started with the goal of nanoscale localization of particles beyond the classical diffraction limit, plasmonic tweezers concept [4],[24] and related auxiliary tools were found to be an efficient approach for tailoring nanoscale mechanical motion with light [25]–[28]. For example, hyperbolic metamaterials and metasurfaces introduced a bunch of new effects, including tractor beams, anti-trapping and several others [29][30],[31]–[34][35]. Other types of auxiliary optomechanical structures include metalenses [36], laser-printed manipulators [37], plasmonic Archimedes spiral lens [38], photonic hooks [39], [40], photonic nanojets [41], [42], [43] and many others.

The last method, to be mentioned in this context, is to shape a particle itself. In a vast majority of cases optomechanical manipulation is performed on spherical particles. Those are typically made of transparent low-index dielectric materials [44], [45] or plasmonic metals [46], [47]. In both cases, however, particles' polarizability is linked to its dipolar response and governs the interaction. Here the balance between gradient forces and radiation pressure dictates the dynamics [48]. The latter can be quite complex owing to nontrivial near fields, created by auxiliary structures [49]–[51]. However, high-index dielectric particles supporting a variety of Mie resonances [52], [53] introduce new interaction channels beyond simple dipolar polarizability terms [54], [55]. For example, coherent interaction between electric and magnetic responses of silicon particles was shown to provide either pulling or pushing forces, depending on system's parameters [56]. Sorting of silicon particles with laser beams was shown in [57]. Core-shell geometries allow designing multipole resonances within a structure [58].

Careful tailoring of multipole interference can provide superior capabilities to control optical forces. Intuitively, a proper combination of multipoles can lead to quite arbitrary far-field scattering pattern. As a result, recoil forces can be flexibly engineered, though electromagnetic interactions in their complete form should be addressed – this is the goal of this report. The influence of conservative and non-conservative forces on particles' dynamics will be studied, and novel optomechanical effects delivered by different electric and magnetic multipoles will be introduced.

The manuscript is organized as follows: after discussing the mathematical formulation of optical forces and linking them to multipolar expansion, conditions for trapping/anti-trapping will be introduced. The detailed studies of force components will follow, unraveling the emergence of new enhanced trapping, anti-trapping and bending phenomena.

# Optical Forces in multipolar description

A typical setup under consideration is depicted in Fig. 1 showing an interaction of an optical beam with a particle. Typically, optical trapping is performed with a focused Gaussian beam, which will be used here. The formalism, however, can be extended to an arbitrary waveform by applying a plane wave expansion (discussed below, e.g. [59] ). Since plane wave scattering on a sphere has closed form analytical solutions (Mie theory), optical force computation relying on a knowledge of self-consistent electromagnetic fields is also computationally efficient. Fig. 1 summarizes the layout, which will be used for the subsequent investigations. A spherical particle is situated at a waist of a linearly polarized Gaussian beam. The forces will be analyzed at the focal plane, transverse to the propagation direction (Fig. 1(b)). The emphasize will be done on far-field scattering diagram analysis (panel (c)), which will be responsible for controlling trapping/anti-trapping conditions.

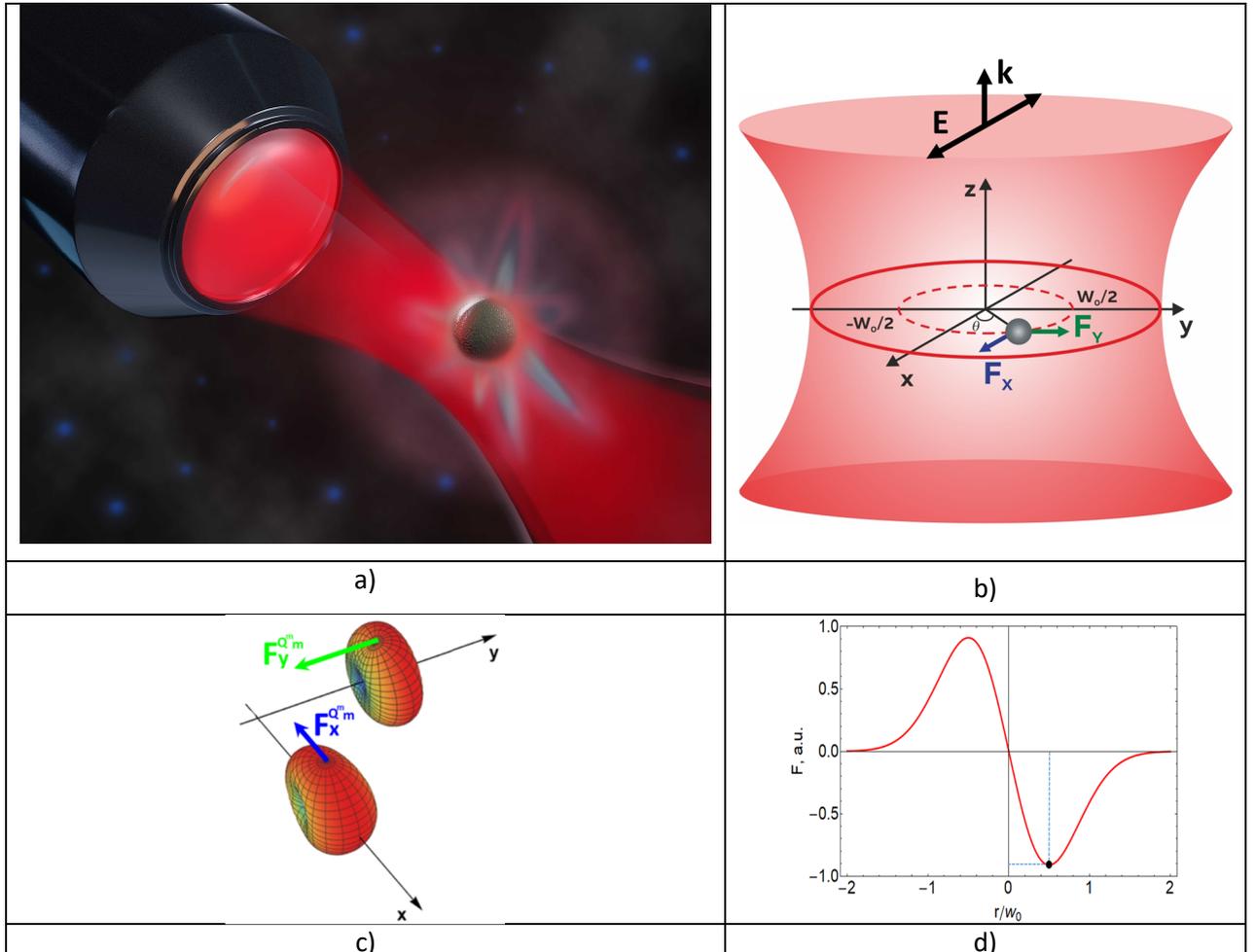

**Fig. 1** (a) Schematics of a particle in an optical trap. (b) A particle in a Gaussian beam with $w_0$ beam waist + coordinate system. (c) An example of scattered field formation by interfering magnetic dipole ($m_y$) and a magnetic quadrupole ($Q^m_{xy}$) for $\theta=0$ and ($Q^m_{yy}$) for $\theta=\pi/2$. The asymmetry of the far-field scattering pattern leads to recoil forces shown by arrows. (d) transverse component of a gradient optical force, acting on a subwavelength dipolar particle. Black point indicates the position, where part of the subsequent calculations are carried out.

Time-averaged optical force ($\langle \mathbf{F} \rangle$) is calculated by integrating Maxwell's stress tensor ($\ddot{\mathbf{T}}$) over a virtual surface (S) enclosing the particle:

$$\langle \mathbf{F} \rangle = \frac{1}{2}\text{Re}\oint_S \ddot{\mathbf{T}} \cdot \mathbf{n} dS ,\qquad(1)$$

where **n** is an outward normal to the enclosing surface. Hereinafter we will use phasor notation, considering $e^{-i\omega t}$ as the time dependence. The medium enclosing the particle is assumed to be vacuum. In this case the stress tensor is given by:

$$\ddot{\mathbf{T}} = \varepsilon_0 \mathbf{E} \otimes \mathbf{E}^* + \mu_0 \mathbf{H} \otimes \mathbf{H}^* - \frac{1}{2}\left(\varepsilon_0 \mathbf{E}\cdot\mathbf{E}^* + \mu_0 \mathbf{H}\cdot\mathbf{H}^*\right)\ddot{\mathbf{I}},\qquad(2)$$

Where $\mathbf{E}=\mathbf{E}_{inc}+\mathbf{E}_{sca}, \mathbf{H}=\mathbf{H}_{inc}+\mathbf{H}_{sca}$ are total self-consistent electromagnetic fields, decomposed into the sum of incident and scattered contributions. Since the shape and size of the enclosing surface does not affect the resulting force in Eq. 1, a sphere with a radius ensuring far-field conditions on its boundary will be chosen. In this case the total force can be decomposed into two contributions:

$$\langle {}^I\mathbf{F}\rangle = \lim_{r\to\infty}\oint_S \text{Re}\left[-\frac{1}{2}\left(\varepsilon_0 \mathbf{E}_{inc}\cdot\mathbf{E}_{sca}^* + \mu_0 \mathbf{H}_{inc}\cdot\mathbf{H}_{sca}^*\right)\mathbf{n}\right]dS$$

$$\langle {}^R\mathbf{F}\rangle = -\frac{1}{4}\lim_{r\to\infty}\oint_S \text{Re}\left[\left(\varepsilon_0 \mathbf{E}_{sca}\cdot\mathbf{E}_{sca}^* + \mu_0 \mathbf{H}_{sca}\cdot\mathbf{H}_{sca}^*\right)\cdot\mathbf{n}\right]dS = -\frac{1}{2}\lim_{r\to\infty}\oint_S \text{Re}\left[\left(\varepsilon_0 \mathbf{E}_{sca}\cdot\mathbf{E}_{sca}^*\right)\cdot\mathbf{n}\right]dS$$

,(3)

where $\langle {}^I\mathbf{F}\rangle$ (interception or extinction force) emerges from the interference between incident and scattered fields, while $\langle {}^R\mathbf{F}\rangle$ (recoil force) depends only on scattering. The later is governed by multipolar interference, as it was shown in [60]–[62] [63] . The recoil force can be directly linked to the scattering diagram assymetry [64]. Multipolar decomposition of the far field is given by [65]:

$$\begin{cases} \mathbf{E}_{sca}(\mathbf{r}) = \frac{k^2}{4\pi\varepsilon_0}\frac{e^{ikr_0}}{r_0}\left[\mathbf{n}\times(\mathbf{p}\times\mathbf{n}) + \frac{1}{c}(\mathbf{m}\times\mathbf{n}) + \frac{ik}{2}\mathbf{n}\times\left[\mathbf{n}\times\left(\mathbf{Q}^e\cdot\mathbf{n}\right)\right] + \frac{ik}{2c}\mathbf{n}\times\left(\mathbf{Q}^m\cdot\mathbf{n}\right)\right], \\ \mathbf{H}_{sca}(\mathbf{r}) = \frac{1}{\mu_0}\mathbf{B}_{sca}(\mathbf{r}) = \frac{1}{Z}\mathbf{n}\times\mathbf{E}_{sca}(\mathbf{r}) \end{cases}\qquad(4)$$

where $Z$ is the wave impedance and the expansion is made up to the quadrupole order. Multipolar moments (electric and magnetic dipoles, electric and magnetic quadrupoles, respectively) are given by:

$$\mathbf{p}=\varepsilon_0\alpha_e\mathbf{E}_{inc},\quad \mathbf{m}=\frac{\alpha_m}{\mu_0}\mathbf{B}_{inc},\quad \mathbf{Q}^e=\varepsilon_0\alpha_{Q^e}\frac{\nabla\mathbf{E}_{inc}+\mathbf{E}_{inc}\nabla}{2},\quad \mathbf{Q}^m=\frac{\alpha_{Q^e}}{\mu_0}\frac{\nabla\mathbf{B}_{inc}+\mathbf{B}_{inc}\nabla}{2}$$

(5)

where the relation $(\nabla\mathbf{A}+\mathbf{A}\nabla)_{ij} = \partial_i A_j + \partial_j A_i$ is used. Polarizabilities are linked to Mie coefficients ($a_1$, $a_2$, $b_1$, and $b_2$ [66]) as follows:

$$\alpha_e = i\frac{6\pi}{k^3}a_1, \quad \alpha_m = i\frac{6\pi}{k^3}b_1, \quad \alpha_{Q^e} = i\frac{40\pi}{k^5}a_2, \quad \alpha_{Q^m} = i\frac{40\pi}{k^5}b_2 \qquad (6)$$

By substituting Eq. 4 into Eq. 3 the following force decomposition is obtained:

$$\begin{cases} \langle F_i \rangle = \langle {}^I F_i \rangle + \langle {}^R F_i \rangle \\ \langle {}^I F_i \rangle = \frac{1}{2}\text{Re}\left[p_j \nabla_i E^*_{inc,j}\right] + \frac{1}{2}\text{Re}\left[m_j \nabla_i B^*_{inc,j}\right] + \frac{1}{4}\text{Re}\left[\left(Q^e\right)_{jk} \nabla_i \nabla_k E^*_{inc,j}\right] + \\ \quad + \frac{1}{4}\text{Re}\left[\left(Q^m\right)_{jk} \nabla_i \nabla_k B^*_{inc,j}\right] \\ \langle {}^R F_i \rangle = -\frac{k^4}{12\pi\varepsilon_0 c}\text{Re}\left[\varepsilon_{ijk} p_j m^*_k\right] - \frac{k^5}{40\pi\varepsilon_0}\text{Im}\left[\left(Q^e\right)_{ij} p^*_j\right] - \frac{k^5}{40\pi\varepsilon_0 c^2}\text{Im}\left[\left(Q^m\right)_{ij} m^*_j\right] - \\ \quad -\frac{k^6}{240\pi\varepsilon_0 c}\text{Re}\left[\varepsilon_{ijk}\left(Q^e\right)_{lj}\left(Q^m\right)^*_{lk}\right], \quad i,j,k \text{ can be x,y or z} \end{cases} \qquad (7)$$

This expression, as Eq. 4, includes contributions up to quadrupole order, which is typically sufficient for describing interactions with submicron particles having refractive indexes about 3-4. Eq. 7 includes self-consistent electromagnetic fields and their spatial derivatives. It is worth noting that these expressions include cross terms, resulting from far-field interference of multipole contributions. Intuitively, their appearance can be understood by revising a simplified scenario. Consider optomechanical interactions with a Kerker particle, where constructive interference of electric and magnetic dipoles leads to the backscattering suppression. As the result of this rather dramatic scattering diagram reshaping (recall that a Gaussian beam can be decomposed into a sum of plane waves), new transverse forces, orthogonal to both gradient contribution and the radiation pressure, can emerge [54].

In order to verify the contribution of different multipoles and justify the series truncation at the quadrupolar term in Eq. 7, scattering efficiency of a silicon 140 nm radius nanoparticle was calculated. From figure 2 we can see that, in this particular case, electric and magnetic octupoles ($O^E$ and $O^M$) have minor contribution to the interaction and they can be safely neglected. It is worth noting that in this investigation we use spherical multipoles as a basis. In this case toroidal moments, which should be explicitly introduced in Cartesian expansion, are included by construction.

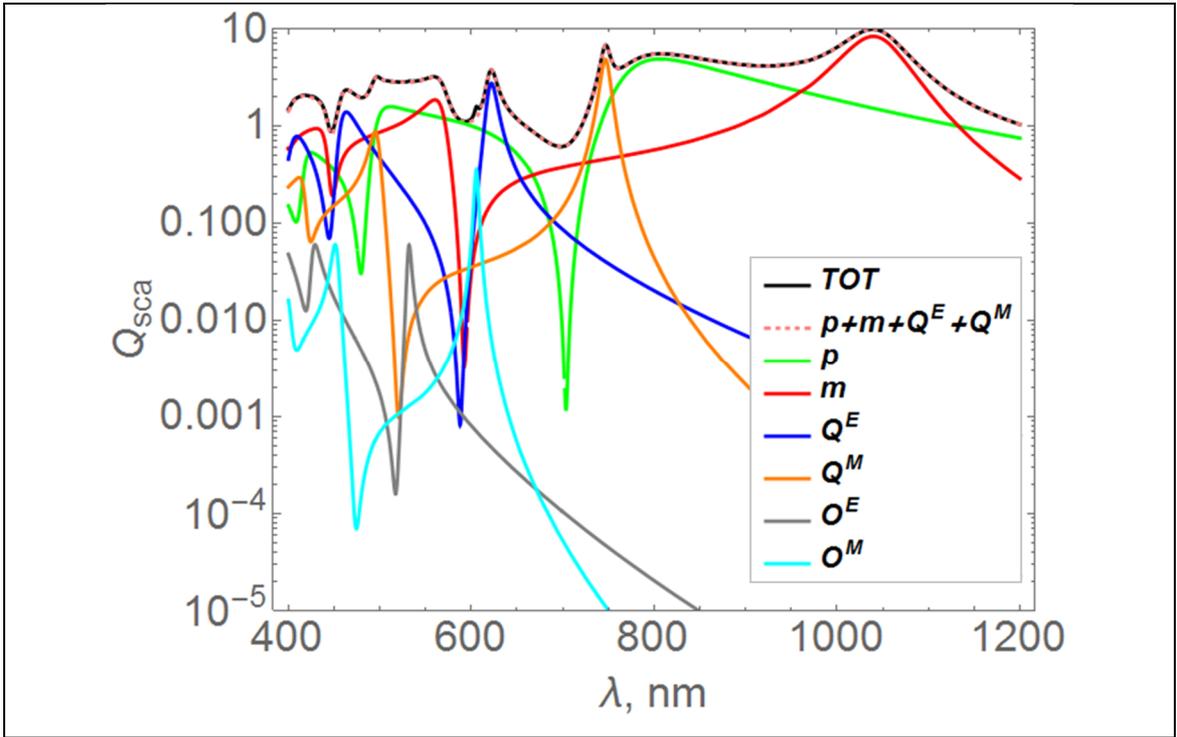

**Fig. 2.** Scattering efficiency of a silicon 140nm radius nanoparticle – total scattering efficiency and multipolar contributions (multipole abbreviations appear in the legend).

While the formalism of Eq. 7 allows calculating optical forces, it can be further modified and brought to the form, where the link between scattering pattern asymmetry and recoil forces is evident. After some mathematical manipulations, the recoil force component is given by:

$$\left\langle {}^R\mathbf{F}\right\rangle = -\frac{1}{c}\oint_S \left(I_p + I_m + I_{Q^e} + I_{Q^m} + c\varepsilon_0 \operatorname{Re}\left[\mathbf{E}_{Q^e}\cdot\mathbf{E}_m^* + \mathbf{E}_{Q^m}\cdot\mathbf{E}_p^*\right]\right)\cdot\mathbf{n}dS -$$
$$-\oint_S \left(\varepsilon_0 \operatorname{Re}\left[\mathbf{E}_p\cdot\mathbf{E}_m^* + \mathbf{E}_{Q^e}\cdot\mathbf{E}_{Q^m}^* + \mathbf{E}_{Q^e}\cdot\mathbf{E}_p^* + \mathbf{E}_{Q^m}\cdot\mathbf{E}_m^*\right]\right)\cdot\mathbf{n}dS \qquad (8)$$

where far-field intensities $I_p, I_m, I_{Q^e}, I_{Q^m}$ are obtained by substituting the electric field of the standalone multipole into the expression $I = \frac{c\varepsilon_0}{2}\mathbf{E}\cdot\mathbf{E}^*$ and all labeled fields correspond to the field scattered from the indicated multipole. Electric fields of multipoles can be either symmetric (do not change their sign when the vector $\mathbf{n}$ is replaced by $-\mathbf{n}$ in (4) for $\mathbf{E}_p, \mathbf{E}_{Q^m}$), or antisymmetric (in the opposite case: $\mathbf{E}_m, \mathbf{E}_{Q^e}$). Consequently, multiplying fields of different symmetry leads to an angular asymmetry in the cross-terms of Eq. 8. Considering this, and due to the fact that single multipoles have a symmetric intensity distribution, the first integral in Eq. 8 vanishes, which leads to a simplified form of the recoil force:

$$\left\langle {}^R\mathbf{F}\right\rangle = -\oint_S \left(\varepsilon_0 \operatorname{Re}\left[\mathbf{E}_p\cdot\mathbf{E}_m^* + \mathbf{E}_{Q^e}\cdot\mathbf{E}_{Q^m}^* + \mathbf{E}_{Q^e}\cdot\mathbf{E}_p^* + \mathbf{E}_{Q^m}\cdot\mathbf{E}_m^*\right]\right)\cdot\mathbf{n}dS \qquad (9)$$

Transverse components of the recoil optical force can be calculated by projecting vectorial force on the Cartesian coordinate system as follows:

$$^{R}F_{x}^{pm} = -\frac{k^{4}}{12\pi\varepsilon_{0}c}\text{Re}\left[p_{y}m_{z}^{*} - p_{z}m_{y}^{*}\right]; \quad ^{R}F_{y}^{pm} = -\frac{k^{4}}{12\pi\varepsilon_{0}c}\text{Re}\left[p_{z}m_{x}^{*} - p_{x}m_{z}^{*}\right]$$

$$^{R}F_{x}^{Q^{e}Q^{m}} = -\frac{k^{6}}{240\pi\varepsilon_{0}c}\text{Re}\begin{bmatrix}(Q^{e})_{xy}(Q^{m})_{xz}^{*} - (Q^{e})_{xz}(Q^{m})_{xy}^{*} + (Q^{e})_{yy}(Q^{m})_{yz}^{*} - (Q^{e})_{yz}(Q^{m})_{yy}^{*} + \\ +(Q^{e})_{zy}(Q^{m})_{zz}^{*} - (Q^{e})_{zz}(Q^{m})_{zy}^{*}\end{bmatrix}$$

$$^{R}F_{y}^{Q^{e}Q^{m}} = -\frac{k^{6}}{240\pi\varepsilon_{0}c}\text{Re}\begin{bmatrix}-(Q^{e})_{xx}(Q^{m})_{xz}^{*} + (Q^{e})_{xz}(Q^{m})_{xx}^{*} - (Q^{e})_{yx}(Q^{m})_{yz}^{*} + (Q^{e})_{yz}(Q^{m})_{yx}^{*} - \\ -(Q^{e})_{zx}(Q^{m})_{zz}^{*} + (Q^{e})_{zz}(Q^{m})_{zx}^{*}\end{bmatrix}$$

$$^{R}F_{x}^{Q^{e}p} = -\frac{k^{5}}{40\pi\varepsilon_{0}}\text{Im}\left[(Q^{e})_{xx}p_{x}^{*} + (Q^{e})_{xy}p_{y}^{*} + (Q^{e})_{xz}p_{z}^{*}\right] \quad (10).$$

$$^{R}F_{y}^{Q^{e}p} = -\frac{k^{5}}{40\pi\varepsilon_{0}}\text{Im}\left[(Q^{e})_{yx}p_{x}^{*} + (Q^{e})_{yy}p_{y}^{*} + (Q^{e})_{yz}p_{z}^{*}\right]$$

$$^{R}F_{x}^{Q^{m}m} = -\frac{k^{5}}{40\pi\varepsilon_{0}}\text{Im}\left[(Q^{m})_{xx}m_{x}^{*} + (Q^{m})_{xy}m_{y}^{*} + (Q^{m})_{xz}m_{z}^{*}\right]$$

$$^{R}F_{y}^{Q^{m}m} = -\frac{k^{5}}{40\pi\varepsilon_{0}}\text{Im}\left[(Q^{m})_{yx}m_{x}^{*} + (Q^{m})_{yy}m_{y}^{*} + (Q^{m})_{yz}m_{z}^{*}\right]$$

Eq. 10 provides explicit expressions underlining the link between the scattered pattern asymmetry and attraction/repulsion recoil forces.

The next step is to analyze the conditions at which, optical forces acting on a small particle, are significantly different from the standard case, where only dipolar polarizability is taken into account. Typical trapping layout appears in Fig. 1. The following parameters are considered: the beam waist $w_0$=5μm (loosely focused beam), electrical field amplitude (at the beam's center) $E_0$=10$^6$[V/m]. The beam is linearly polarized along x-axis, the CW laser wavelength is the subject to the forthcoming parametric study. Silicon nanoparticle is placed r = $w_0$/2 (black point in Fig. 1(b)), while azimuthal angle θ is a variable.

Angular spectral decomposition of a Gaussian beam is used to calculate the scattering pattern [67]. Fourier series for electric and magnetic fields are given as follows:

$$\mathbf{E}_{inc}(x,y,z) = \int\int_{-\infty}^{+\infty}\hat{\mathbf{E}}(k_x,k_y;0)e^{i(k_x x + k_y y + k_z z)}dk_x dk_y$$

$$\mathbf{H}_{inc}(x,y,z) = \int\int_{-\infty}^{+\infty}\hat{\mathbf{H}}(k_x,k_y;0)e^{i(k_x x + k_y y + k_z z)}dk_x dk_y$$

(11)

where $\hat{\mathbf{E}}(k_x,k_y;0)$ and $\hat{\mathbf{H}}(k_x,k_y;0)$ are the field amplitudes at the beam waist and $k_i, i=x,y,z$ are the wave vector components in the Cartesian coordinate system. For x-axis linearly polarized fundamental mode, the electric field $\mathbf{E}_{inc} = (E_x, 0, E_z)$ can be written in as:

$$\hat{E}_x(k_x, k_y; 0) = E_0 \frac{w_0^2}{4\pi} e^{-(k_x^2+k_y^2)\frac{w_0^2}{4}}, \qquad (12)$$

where the longitudinal z-component is insignificant (in more complex scenarios it though plays an important role [68]). Using Faraday's law, the magnetic field components are straightforwardly derived. Finally, the incident fields are given by:

$$\mathbf{E}_{inc}(x,y,z) = \int\!\!\int_{-\infty}^{+\infty} \hat{E}_x(k_x,k_y;0) \frac{1}{k_z}[k_z \mathbf{n}_x - k_x \mathbf{n}_z] e^{i(k_x x + k_y y + k_z z)} dk_x dk_y$$

$$\mathbf{H}_{inc}(x,y,z) = Z^{-1}\int\!\!\int_{-\infty}^{+\infty} \hat{E}_x(k_x,k_y;0) \frac{1}{kk_z}\left[-k_x k_y \mathbf{n}_x + (k_x^2+k_z^2)\mathbf{n}_y - k_y k_z \mathbf{n}_z\right] e^{i(k_x x + k_y y + k_z z)} dk_x dk_y \qquad (13)$$

where $\mathbf{n}_i, i=x,y,z$ are unitary vectors.

## Forces analysis

### Anti-trapping conditions

Given the above-mentioned plane wave decomposition of the beam, the Mie problem is solved and self-consistent electromagnetic fields are calculated. Then, those fields are introduced within Eq. 7 and optical forces are calculated. Fig. 3 shows the results of the parametric study, where transverse forces ($F_x$ ($\theta=0$) and $F_y$ ($\theta=\pi/2$)) are investigated as the function of the particle radius and the illumination wavelength. The colored areas on the map correspond to the anti-trapping regime, while the greyscale parts show conditions where typical trapping takes place (recall the particle's position and coordinate system, defined in Fig. 1). A 140nm radius silicon (material dispersion [69]) particle is used in the subsequent studies.

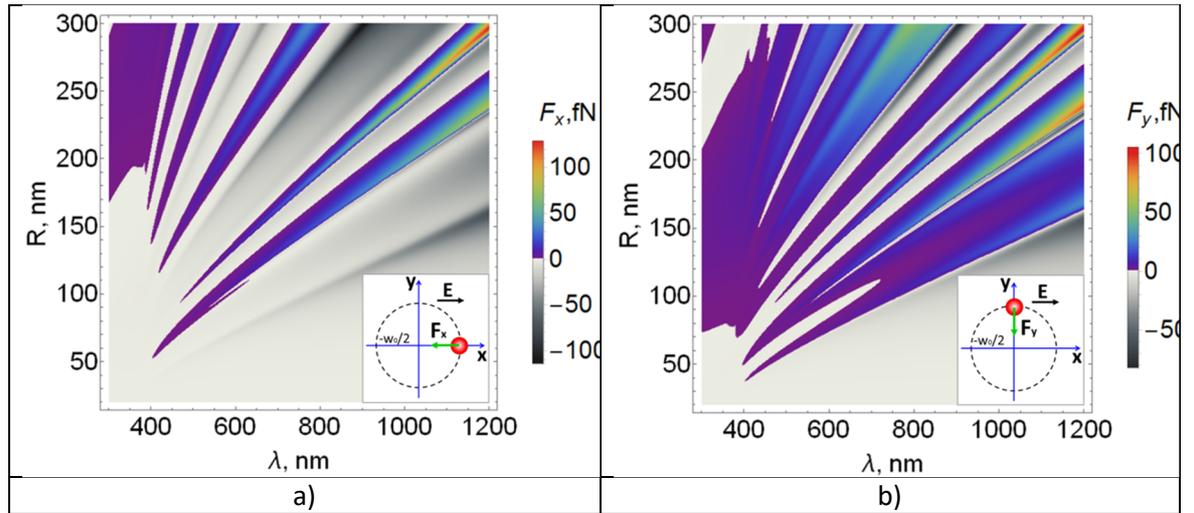

**Fig. 3**. Forces color maps, as the function of particle's radius and the illumination wavelength. (a) $F_x$, $\theta=0$ (b) $F_y$, $\theta=\pi/2$. Color regions correspond to anti-trapping regime. Grey-scale areas correspond to conditions for an attraction force. The insets show the positions of the particle in the beam waist.

The examination of the results from Fig. 3 shows the emergence of anti-trapping regime, which is quite unusual to typical optical trapping scenarios, when low-contrast particles are in use. It is also worth noting that panels (a) and (b) are not entirely symmetric – this is the result of the well-defined polarization of the incident beam, which affects the interference terms in Eq. 10 quite differently. In other words, the asymmetry of $F_x$ and $F_y$ is a direct consequence of the recoil forces radial component contribution. To prove this statement, we will consider the angular dependences of the interception and recoil forces for both, the radial and azimuthal components.

**Anti-trapping - angular dependence**

As it was previously mentioned, linear polarization breaks the rotational symmetry of the problem and, hence, the optical force has an angular dependence. Fig. 4 summarizes this behavior, where we have made the transition between Cartesian and Cylindrical coordinate systems. The surfaces have color gradients along θ and underline the non-uniform angular dependence. In overall, the optical force is not radial (typical conservative gradient force), which will lead to nontrivial particle's trajectories, as it will be discussed later. This angular dependence of the radial force is especially pronounced for shorter wavelengths, where higher-order multipoles play a role (Fig. 2).

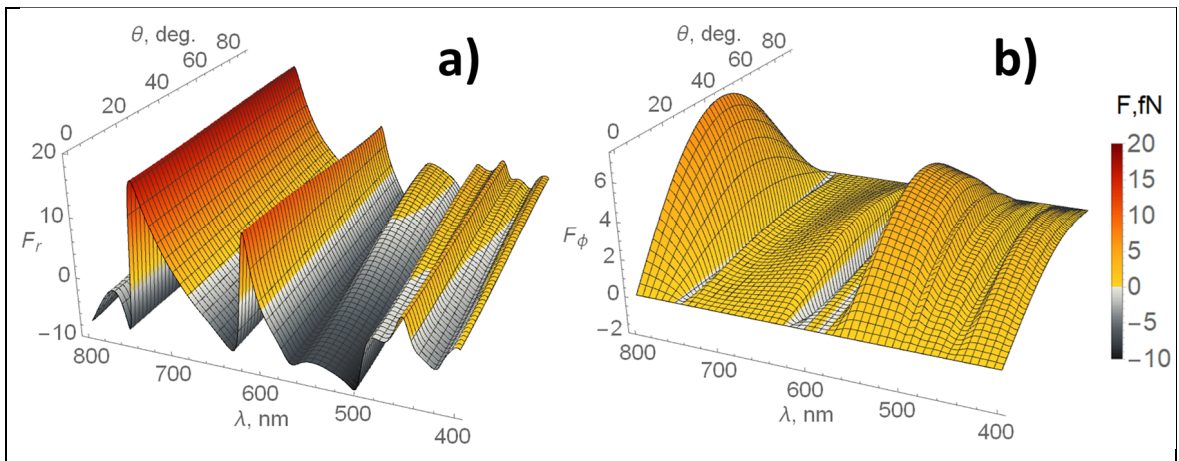

**Fig. 4**. Force components color surfaces. a) $F_r$ and b) $F_\phi$ as a function of the illumination wavelength and angular position of the particle with respect to the beam polarization (Fig. 1). For panel a) color regions correspond to the anti-trapping regime while grey-scale areas corresponds to an attractive force. For panel b) color regions correspond to the azimuthal component of the force that deflects the particle trajectory towards the y-axis; grey-scale areas – towards the x-axis (see figure S3 in supplementary information).

The sign of the radial component of the force dictates whenever the regime is trapping or anti-trapping. As we will next show, the angular asymmetry of radial component, shown in Fig. 4(a), is a direct consequence of the interplaying interception and recoil components. The azimuthal component, on the other hand, has a symmetric angular distribution (relative to π/4 angle) and is responsible for the deviation of particle's trajectory from a radial motion. It drives the particle towards the X or Y axis depending on wavelength (see figure S4 in supplementary information).

Fig. 5 (a) shows the interception force in the focal plane, containing the beam waist. Interception forces are dominated by a gradient derived radial component not depending on the angle [70]–[72]. Similar considerations were made by observing multipolar particle in a Bessel beam [73]. It is worth noting that the conservative nature of interception force in beam waist is determined exclusively by the structure of the electromagnetic field. A flat wave front in the waist plane causes the light pressure force acting on individual multipoles to be directed strictly along the beam propagation direction with no contribution to the transversal force. This effect does not depend qualitatively on the size and material of the particle, provided that the latter is spherical and isotropic.

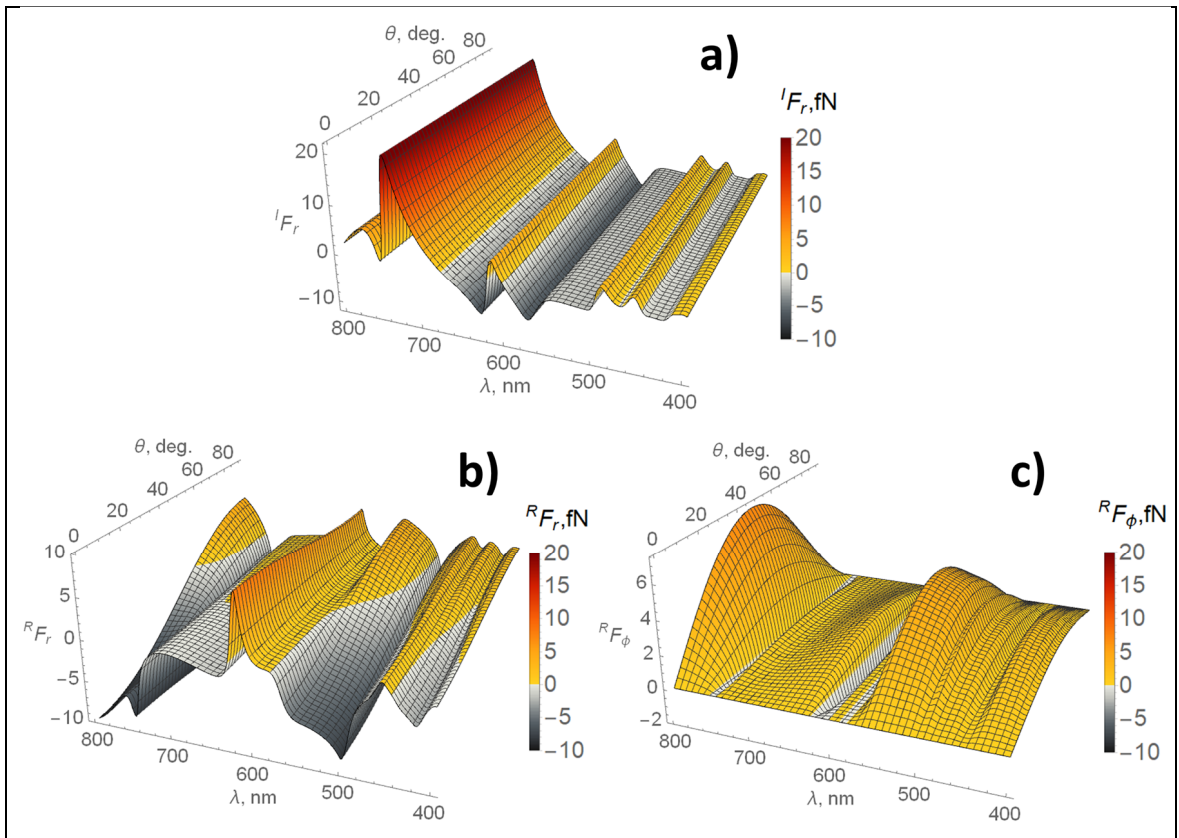

**Fig. 5**. Color surfaces of radial and azimuthal force components, as a function of the illumination wavelength and angular position of the particle. (a) Interception force; (b), (c) recoil force. Color regions for (a) and (b) correspond to anti-trapping regime; grey-scale areas correspond to the conditions for an attractive force. For panel (c) color regions correspond to the azimuthal component of the recoil force that deflects the particle trajectory towards the y-axis while grey-scale areas deflect towards the x-axis (see figure S3 in supplementary information). Note that for interception force, panel a) shows only the radial component, since the azimuthal component is zero.

In optical tweezers the intensity gradient value is negative at the point in space where the particle is localized in our study and, in order to obtain an anti-trapping interception force in a dipole, a negative real part of the polarizability is needed. This is possible for plasmonic materials like silver but not for a low contrast nanoparticle. However, it should be noted that, for the material considered in the present study, when more multipoles are excited, the sign of the interception force, determined by the sign of the corresponding standalone multipole polarizability (Eq. (6)) may be negative, resulting in the existence of a repulsive force for non-metallic materials. (see details in supporting information Fig. S1).

The recoil force can also be divided into radial and azimuthal components [62], [63]. Fig. 5(b) shows that the angular dependence of the total radial force (Fig. 4) is completely determined by the recoil force. The latter depends on the phase difference between pairs of interacting multipoles, as it might be seen from Eq. (9). Comparing Figs. 4(b) and 5(c), we note that the recoil force is responsible for the boost of the azimuthal component.

**Multipole analysis of optical forces**

While the general behavior of interception and recoil forces was analyzed in the previous section, the contribution of multipoles into this behavior will be analyzed next. Figs. 6 (a and b) show the radial force spectra, underlining the interception and recoil force term contributions. The areas of interest are those, where the force is positive (anti-trapping regime).

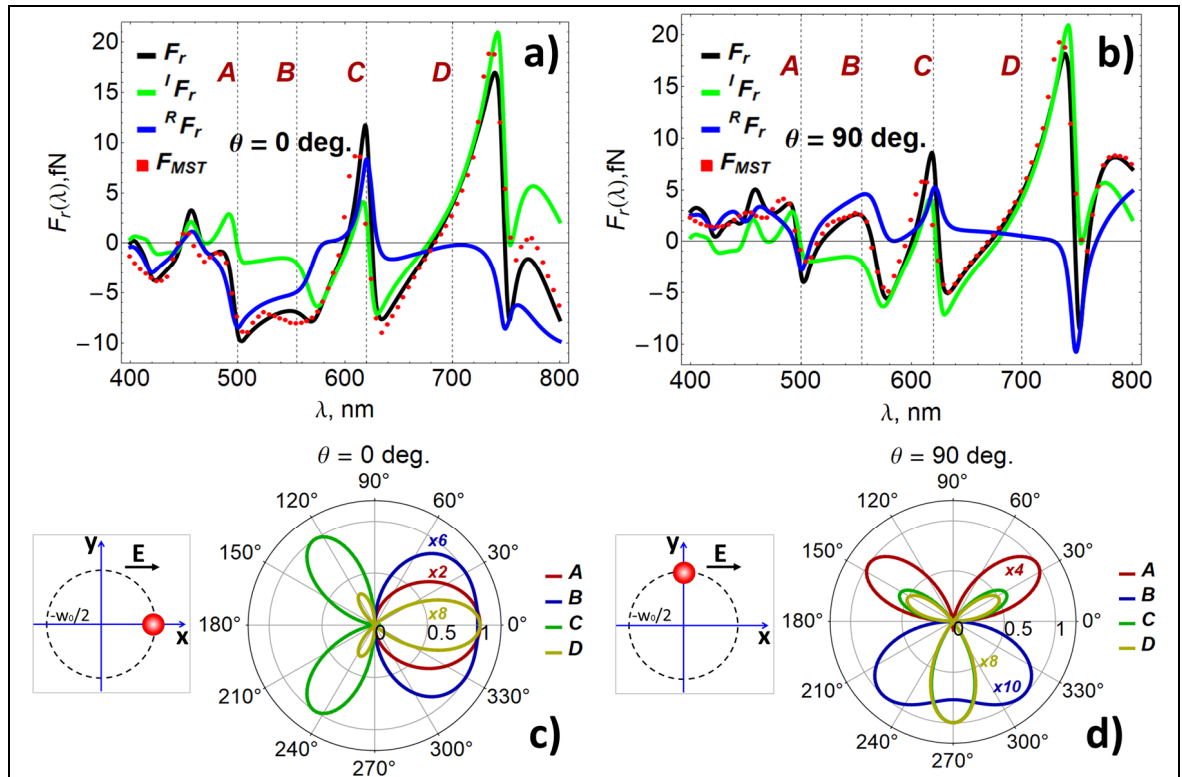

**Fig. 6.** Contribution of the radial interception (green) and recoil components (blue) to the total transversal force (black), acting on a silicon particle (R= 140 nm). (a) $\theta=0$, (b) $\theta=\pi/2$. The red dots show the net force obtained by integrating Maxwell's stress tensor (Eq. 1). Note how interception forces are angle independent and how both, negative and positive values, are possible. In (c, d) the asymmetry of the resulting scattering pattern in the beam waist is plotted. The asymmetry is obtained by subtracting the intensity angular dependence in the X-axis negative direction from the intensity angular dependence in the X-axis positive direction (c) and Y axis (d). The scattering asymmetry determines the direction and magnitude of the recoil force. The insets show the positions of the particle in the beam waist.

To get a closer inspection of the behavior demonstrated in Fig. 6, multipolar components will be investigated in detail next. Several characteristic points have been selected (highlighted in Figs. 6 (a and b) with vertical gray dashed lines):

- ***point "A"*** ($\lambda = 500\,nm$), in the x axis direction ($\theta$ = 0) and y axis direction ($\theta$ = π / 2), optical **trapping** is observed (the value of net force is negative (see Fig. 6 a, b)), while the main contribution to the force is made by the recoil term (the value of the interception term can be neglected);

- ***point "B"*** ($\lambda = 555\,nm$), optical **bending** is obtained. At $\theta$ = 0, the trapping effect is observed, while at $\theta$ = π / 2, the value of net force changes sign to "plus" and the particle is pushed out of the beam. The sign of the net force in this case is determined by the recoil term (since the value of the interception term is always negative and does not depend on the angle $\theta$);

- ***point "C"*** ($\lambda = 620\,nm$), for any angle $\theta$, the **anti-trapping** effect is observed, while the contributions of the interception term and interference multipole terms to the value of net force are comparable;

- ***point "D"*** ($\lambda = 700\,nm$), there is also an angle-independent **anti-trapping** effect in this case defined by the conservative interception force component. The influence of recoil term is insignificant.

Figs. 6 (c) and (d) summarize the far-field patterns for conditions of points "A", "B", "C", and "D". Relationships between the angular asymmetry of the radiation patterns in the plane of the beam waist and the recoil force, arising from the interaction between the multipoles, can be seen. The asymmetry is the result of the fact that the integrand in Eq. (9) can have both negative and positive values. The construction of an asymmetric radiation pattern of interacting multipoles is a good heuristic method for determining the direction of the recoil optical force vector, while Eq. (9) is an alternative way to calculate its numerical value.

The nature of the optical forces described above is demonstrated on vector field maps, showing the spatial distribution of the net force **F** (Fig. 7). For "C" and "D", the force is predominantly radial while, the forces calculated for points "A" and "B", include a significant azimuthal component (see angular distribution of azimuthal components of the net force in Supporting Information Fig. S4a). The radial component in these cases has a pronounced angular dependence and, for point B, the force changes sign allowing for a bending in the particle trajectory from a horizontal direction to a vertical route. The change in the sign of the total force is the result of the interaction between interception and recoil terms (see Figs. 5 (a) and (b)) (see also vector maps of radial and azimuthal components of the net force in Supporting Information Fig. S2 and S3).

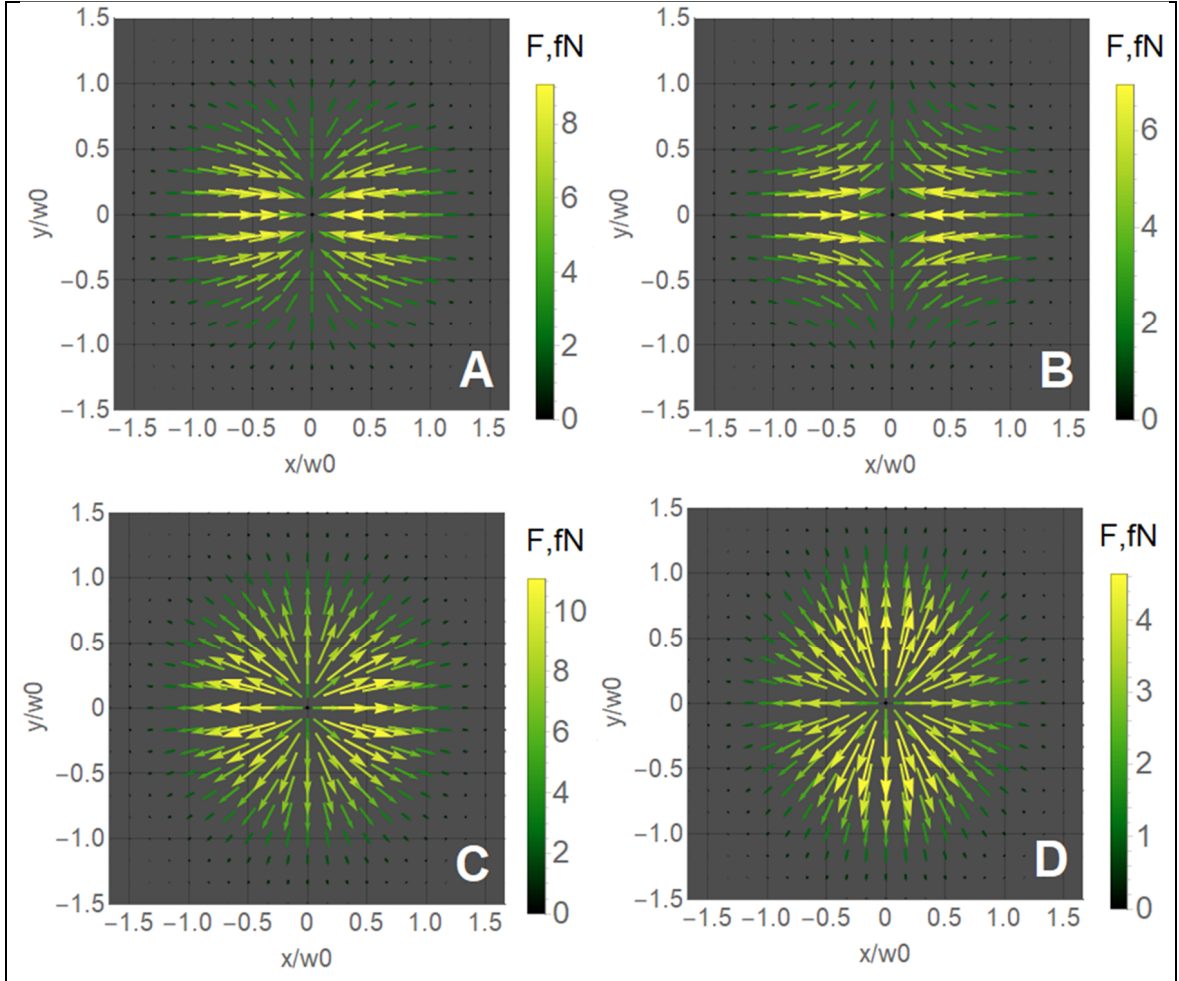

**Fig. 7.** Spatial distribution of the net force in the XY plane for points "A", "B", "C" and "D".

Next, we will investigate contributions of individual multipole components to the resulting optical force. Fig. 8 shows curves of different components from Eq. 7. As can be seen from the comparison of Figs. 8(a) and (b), the anti-trapping at point "D" is determined by the magnetic components of the optical force $^I F^m$ and $^I F^{Q^m}$.

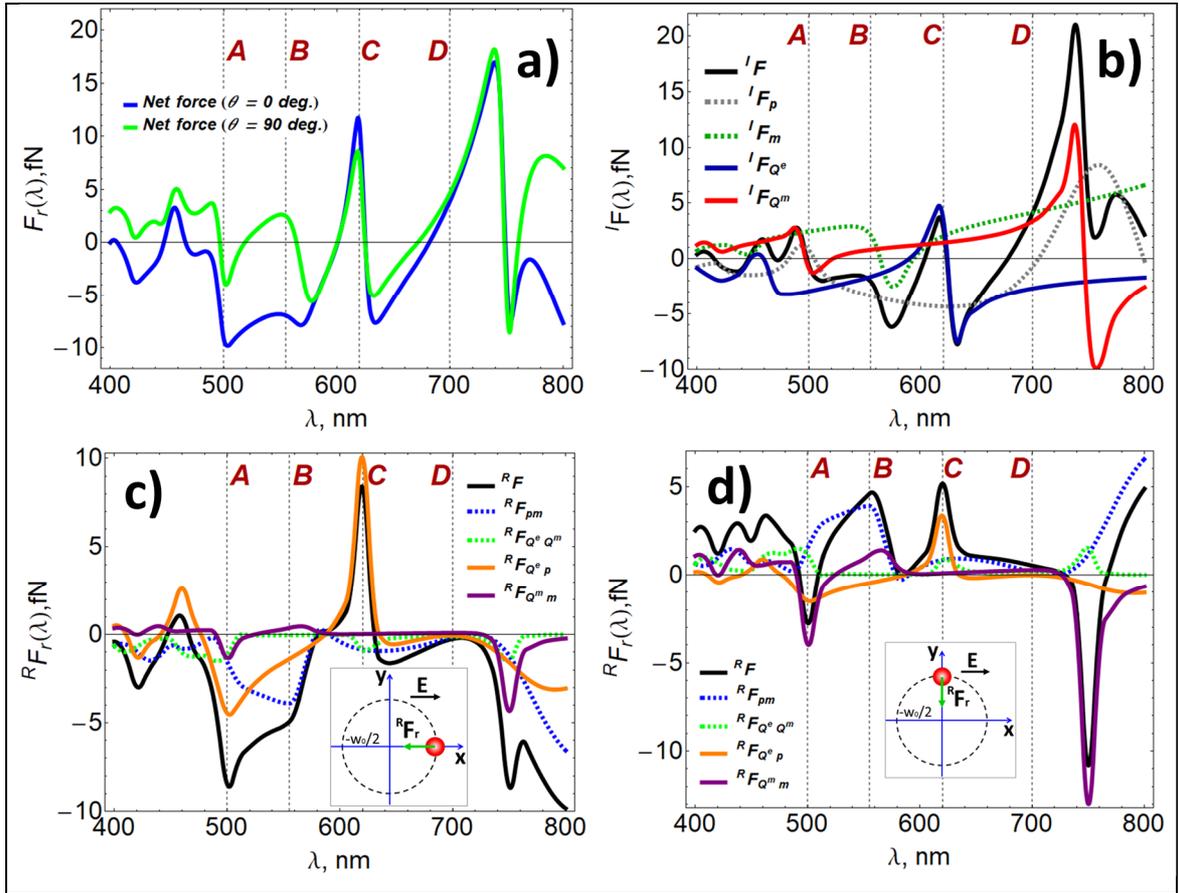

**Fig. 8**. Radial component of the total force (a) and multipole expansion of (b) interception and (c), (d) recoil optical force components, acting on 140nm silicone particle. Total force and multipolar contributions appear in legends. (c) $F_r$, $\theta=0$ (d) $F_r$, $\theta=\pi/2$.

Note that for the considered nanoparticle in the spectral window from 625 to 730 nm, the interception term also makes the main contribution to the net force. For example, at a wavelength of 630 nm, the trapping effect is observed, while the main contribution to this phenomenon is made by the electrical multipole terms of the optical forces $^{I}F^{p}$ and $^{I}F^{Q^e}$. Recall that in this case, when we consider the terms of the force corresponding to the interaction of individual multipoles with an external electromagnetic field, its sign is determined by the phase of these multipole moments.

On the contrary, the trapping effect observed for point "A" is determined only by the interference terms $^{R}F^{Q^e p}$ and $^{R}F^{Q^m m}$ (see Fig. 8 c and d). As shown above, such a force effect can be associated with the asymmetry of the scattered radiation pattern.

Point "B" corresponds to an intermediate scenario where the contribution of both, interception and recoil terms, must be considered. But since the magnitude and sign of the interception term does not depend on the angle $\theta$, the change on direction of the net force and the bending effect is determined by the dipole interference term $^{R}F^{pm}$. Point "C" also has contributions from interception and recoil terms but, in this case, is the dipole-quadrupole interference term $^{R}F^{Q^e p}$ the one leading the force direction. It should be noted that this transverse anti-trapping effect, obtained because of the two electric multipoles interference, is reported for the first time. In a Gaussian beam, the anti-trapping effect was previously obtained only for interfering electric and magnetic dipoles [46]. Obtaining an anti-trapping regime on dipole-quadrupole interference of the same nature moments is inherent for high-index particles only.

We stress that the main difference between cases "C" and "B" is that, in the former case, since $^R F^{Q^e p}$ does not change its sign depending on the angle $\theta$, the anti-trapping effect is always realized while, in the last case, since the the angular dependence of $^R F^{pm}$ is accompanied by a change on sign (Fig. 8 c and d) a bending effect is induced with a trapping mode at $\theta = 0$, and an anti-trapping mode at $\theta = \pi / 2$.

In addition, we should pay attention to the following. While the constructive interference between the terms in Eq. 7 has quite a minor impact on the force behavior, the destructive phenomenon does take place and results in a complete nulling of the overall force. This happens even though there is a moderately high gradient of the field intensity. The total optical force is equal to zero only for a set of wavelengths and at $\theta = 0$ and at $\theta = \pi / 2$ (Fig. 8a). There are also areas in which the net force approaches zero for any angle, although strictly not equal to zero. These are the regions with a wavelength of 600, 625 and 750 nm (see the spectral-angular dependences of the contours where net, interception and recoil forces are zero in Supporting Information Fig. S5.).

## Conclusion

The impact of higher-order multipoles on optomechanical interactions between focused laser beams and high refractive index particles has been investigated. It was shown that a proper balance between multipolar contributions allows controlling direction of optical forces, switching between trapping, anti-trapping and bending regimes almost on demand. In particular, it was shown that quadrupole moments are responsible for achieving anti-trapping behavior. Furthermore, the interception force acting on a particle with a high refractive index can change sign depending on the incident light wavelength. For example, the transverse anti-trapping regime, governed by conservative interception forces (recoil forces in this case can be neglected), corresponds to the magnetic dipole and quadrupole moments of a particle. This effect is atypical for the classical optical tweezers, where the trapping regime is usually implemented. A transverse anti-trapping regime can also emerge in case on interplaying electrical modes only, where the recoil force dominates. Moreover, the very special regime of bending – particle's motion in a curved trajectory could be also realized via different angle-dependencies of the recoil and interception forces. All those regimes can be viewed in the light of far-field interference between higher-order multipoles, where asymmetry factor plays the key role.

Introduction of multipolar degrees of freedom into optomechanical interactions enlarges the capabilities to motion control at the nanoscale, opening a room of opportunities to new possible applications in optics, biology, medicine and lab-on-a-chip platforms.

## Acknowledgments:

The authors are grateful to Natalia Kostina for a productive discussion of the scientific results presented in the article. The force calculations have been supported by the Russian Science Foundation (Grant No. 21-12-00279). MIM acknowledges financial support from the Spanish Ministerio de Ciencia e Innovación (MELODIA PGC2018-095777-B-C-22), the UAM-CAM project (SI1/PJI/2019-00052) and the "María de Maeztu" Programme for Units of Excellence in R&D (CEX2018-000805-M). P.G. acknowledges the support from ERC StG "In Motion" (802279).

# Multipole Engineering of Attractive-Repulsive and Bending Optical Forces

Denis A. Kislov, Egor A. Gurvitz, Alexander A. Pavlov, Manuel I. Marqués, Pavel Ginzburg and Alexander S. Shalin

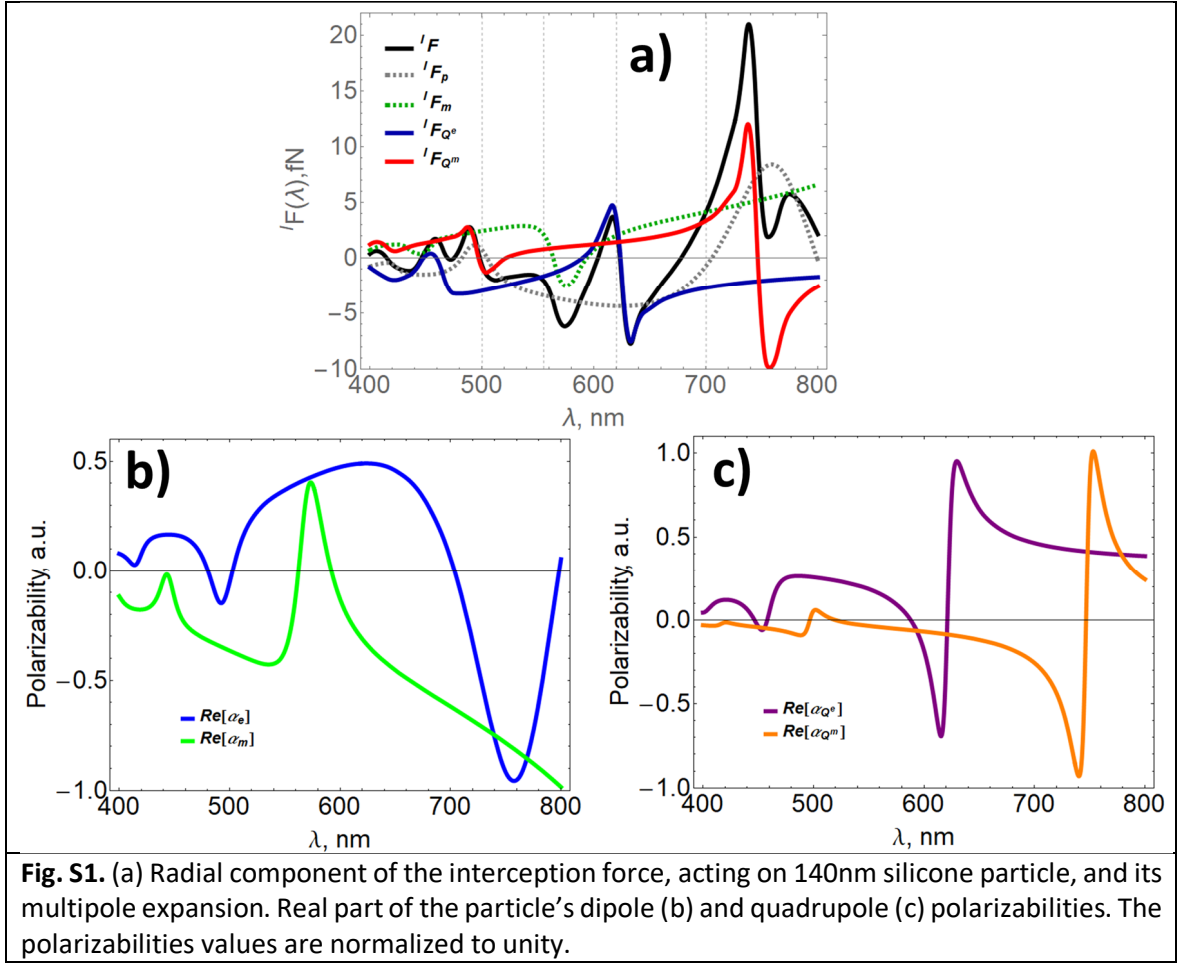

**Fig. S1.** (a) Radial component of the interception force, acting on 140nm silicone particle, and its multipole expansion. Real part of the particle's dipole (b) and quadrupole (c) polarizabilities. The polarizabilities values are normalized to unity.

The sign of the interception force multipole terms is uniquely determined by the sign of the real part of the corresponding multipole polarizability. This is confirmed by comparing the spectral dependences of the dipole and quadrupole polarizabilities real parts Fig. S1 b) and c) with the dependences of the interception force multipole terms in Fig. S1. a). The change in the sign of the polarizability is associated with a change in the spatial orientation of the induced multipole moment.

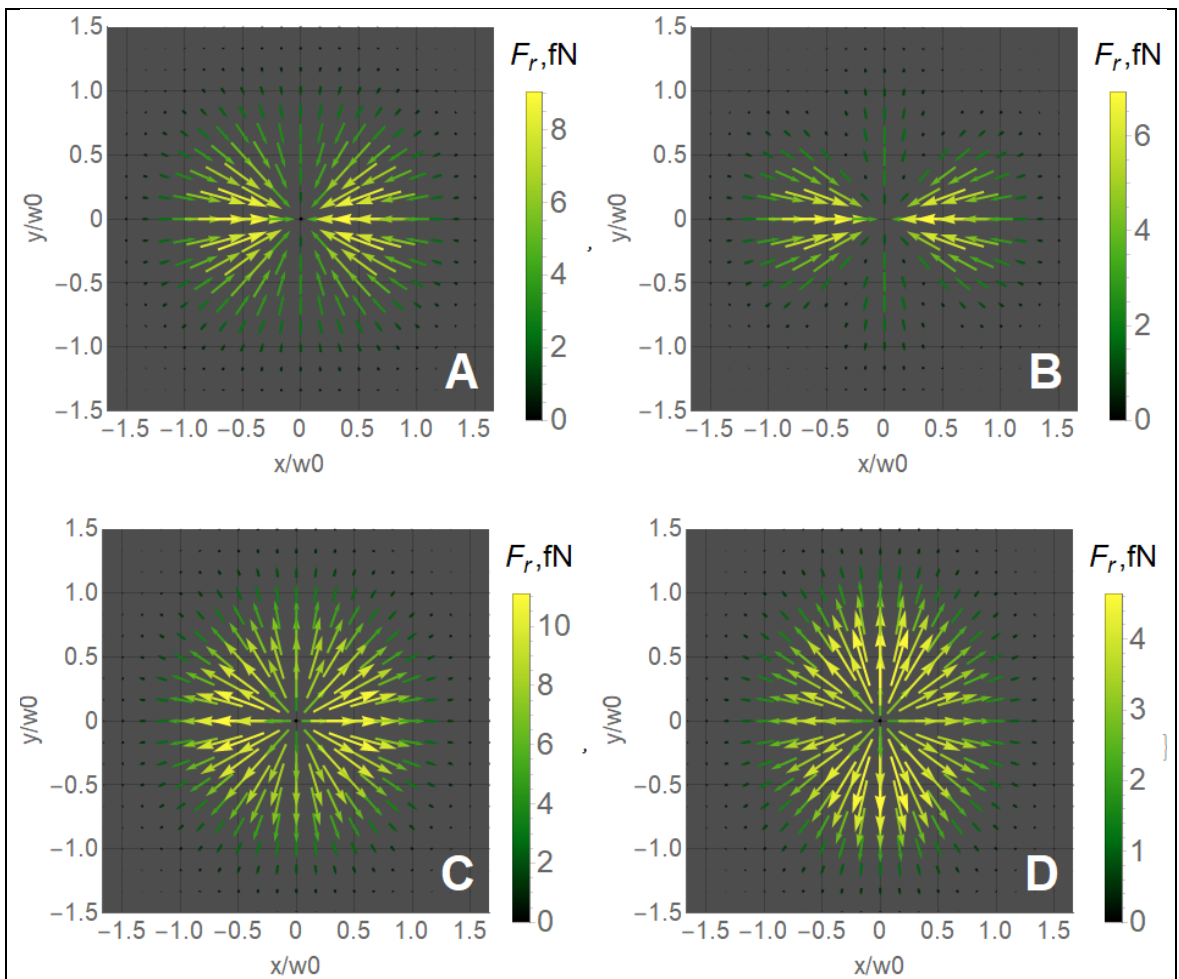

**Fig. S2.** Maps of the spatial distribution radial component of the net force vector fields in the XY plane for points "A", "B", "C" and "D" with the corresponding parameters.

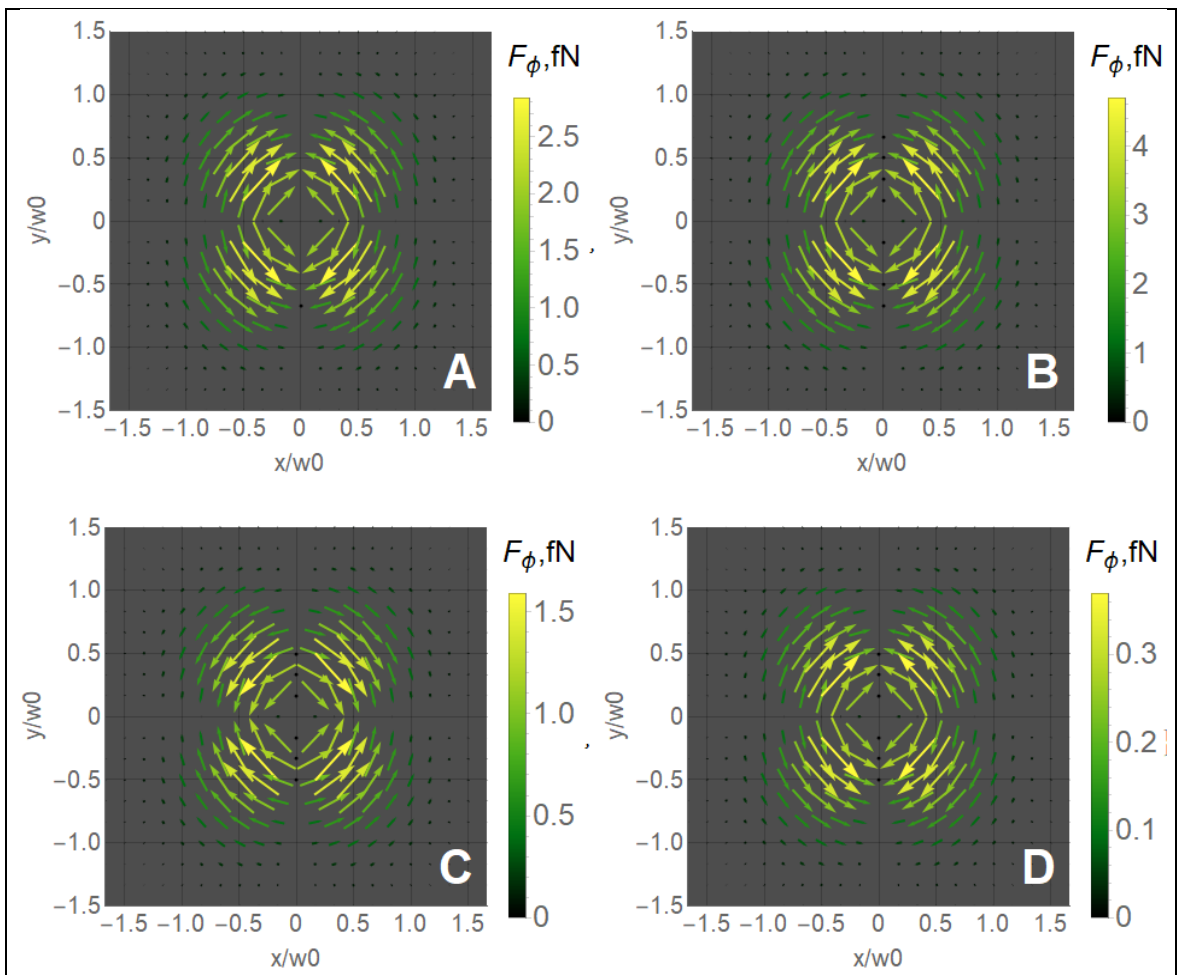

**Fig. S3.** Maps of the spatial distribution azimuthal component of the net force vector fields in the XY plane for points "A", "B", "C" and "D" with the corresponding parameters.

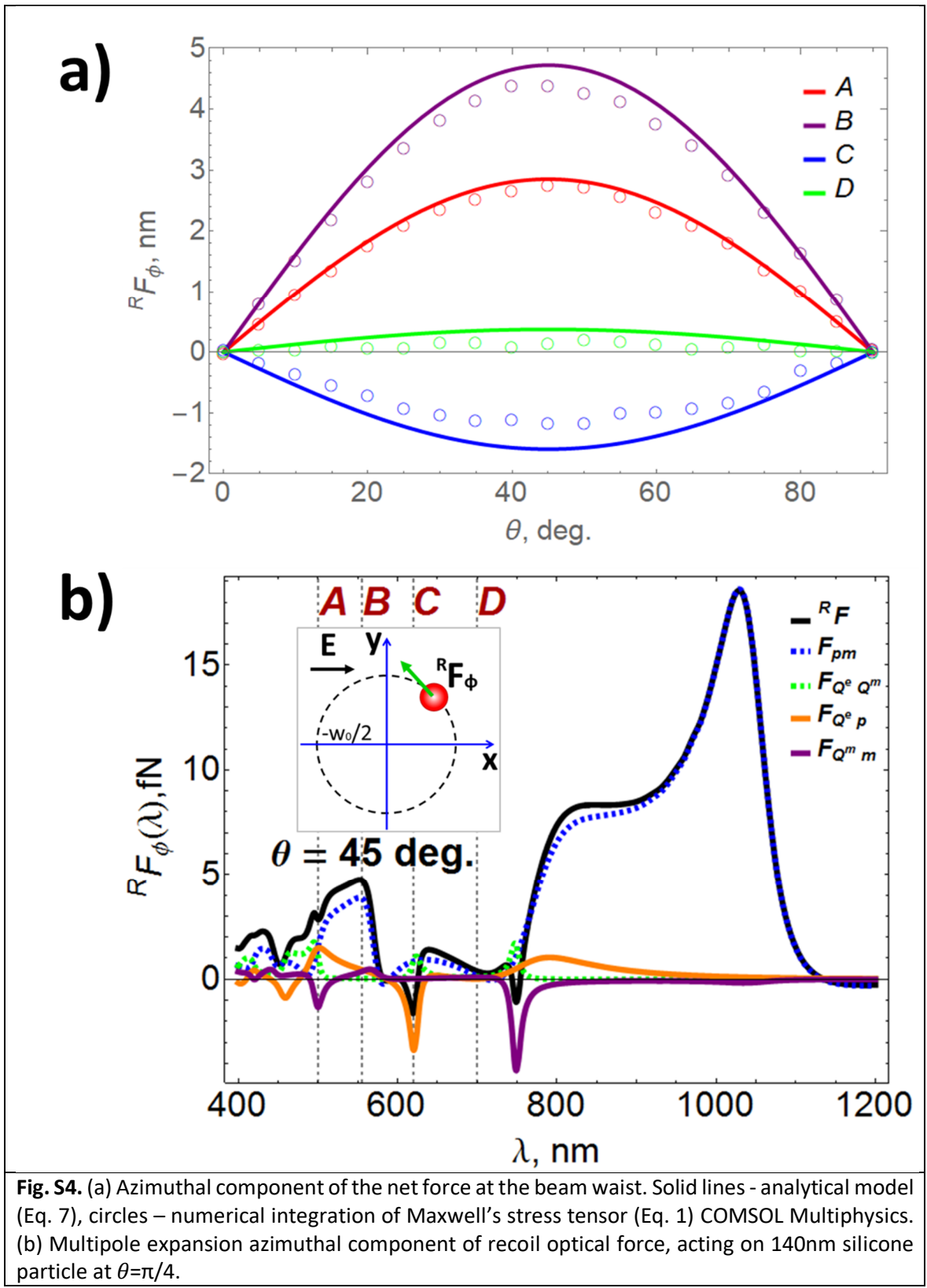

**Fig. S4.** (a) Azimuthal component of the net force at the beam waist. Solid lines - analytical model (Eq. 7), circles – numerical integration of Maxwell's stress tensor (Eq. 1) COMSOL Multiphysics. (b) Multipole expansion azimuthal component of recoil optical force, acting on 140nm silicone particle at $\theta=\pi/4$.

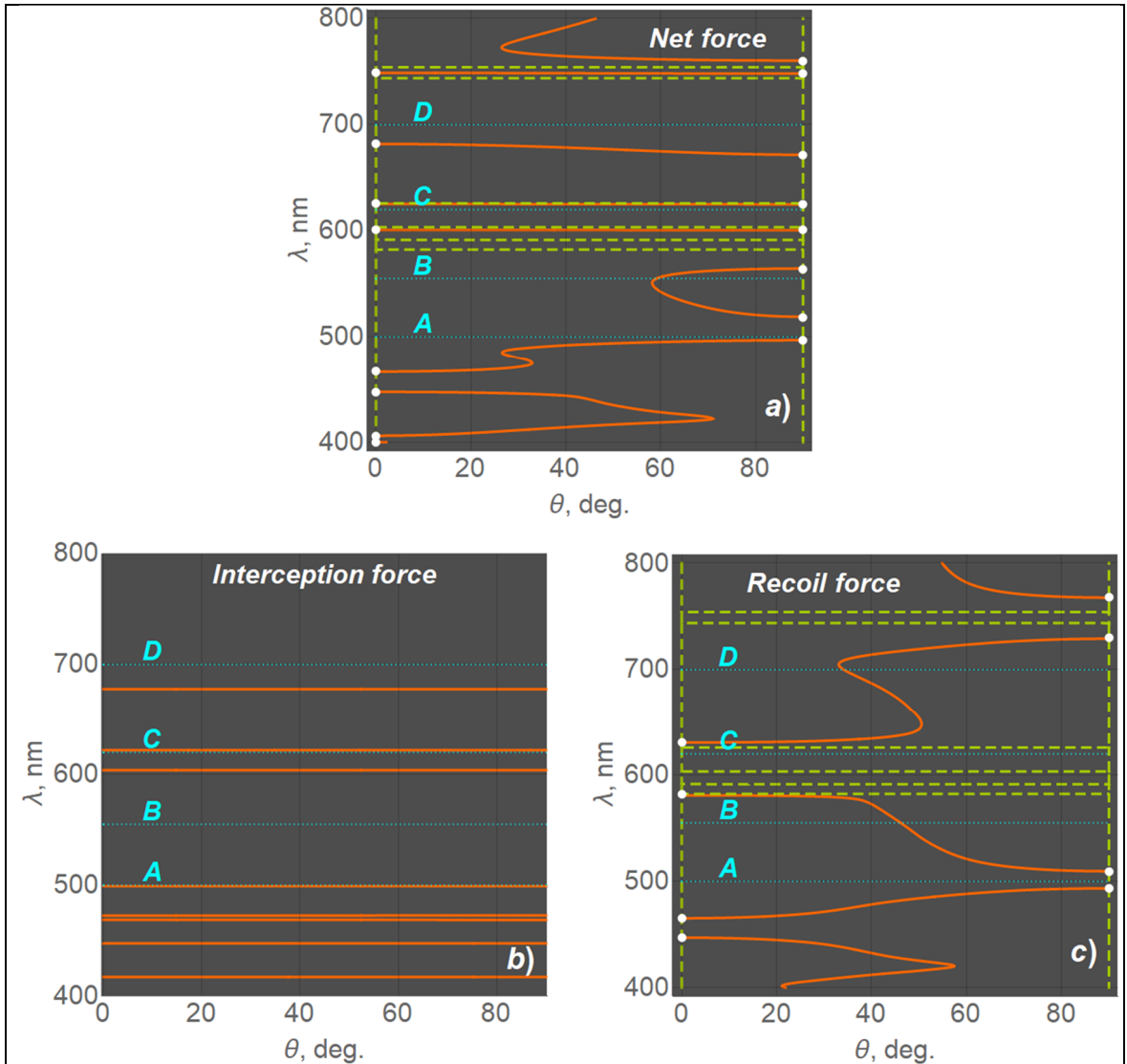

**Fig. S5.** Contours showing the zero values of the radial (orange curves) and azimuthal (green curves) components of the net optical force. White circles mark points where both components are equal to zero simultaneously.